*Title*: **Clusters of calcium release channels harness the Ising phase transition to confine their elementary intracellular signals**

*Short title:* **Ising model approximates Ca sparks termination**

Authors: Anna V. Maltsev[a,1,2], Victor A. Maltsev[b,1], Michael D. Stern[b,1,3]

[1]All authors contributed equally to this work.

*Affiliations:*

[a]Department of Mathematics, University of Bristol, Bristol, United Kingdom

[b]Laboratory of Cardiovascular Science, NIA/NIH, Baltimore, MD, USA

[2]Present address: School of Mathematical Sciences, Queen Mary University of London, London E1 4NS, UK

[3]*Corresponding Author*:

Michael D. Stern,

National Institute on Aging, NIH

Biomedical Research Center

251 Bayview Blvd. Suite 100

Baltimore, MD 21224-6825

Telephone: 410-558-8097

Email: SternMi@mail.nih.gov

**Abstract**:

Intracellular Ca signals represent a universal mechanism of cell function. Messages carried by Ca are local, rapid, and powerful enough to be delivered over the thermal noise. A higher signal to noise ratio is achieved by a cooperative action of Ca release channels such as IP3 receptors or ryanodine receptors arranged in clusters (release units) containing a few to several hundred release channels. The channels synchronize their openings via Ca-induced-Ca-release, generating high-amplitude local Ca signals known as puffs in neurons and sparks in muscle cells. Despite positive feedback nature of the activation, Ca signals are strictly confined in time and space by an unexplained termination mechanism. Here we show that the collective transition of release channels from an open to a closed state is identical to the phase transition associated with the reversal of magnetic field in an Ising ferromagnet. Our new simple quantitative criterion closely predicts the Ca store depletion level required for spark termination for each cluster size. We further formulate exact requirements that a cluster of release channels should satisfy in any cell type for our mapping to the Ising model and the associated formula to remain valid. Thus, we describe deterministically the behavior of a system on a coarser scale (release unit) which is random on a finer scale (release channels), bridging the gap between scales. Our results provide the first exact mapping of a nanoscale biological signaling model to an interacting particle system in statistical physics, making the extensive mathematical apparatus available to quantitative biology.

**Significance Statement:**

Living organisms are built on a hierarchy of levels, starting from macromolecules and clusters of molecules, to organelles, cells, tissues, and with interacting organs finally forming the entire organism. At the lowest of these levels, life depends on individual molecules synchronizing their states to generate robust intracellular signals over the thermal noise. Here we approach the problem via statistical mechanics to describe quantitatively and deterministically this first emerging level of life. We discover that the synchronization that corresponds to termination of local Ca signals generated by clusters of Ca release channels is governed by the same equations as the phase transition associated with the reversal of magnetic field in a classical Ising ferromagnet.

**Introduction**

Clusters of release channels play a fundamental role in biological systems as elementary signal generating, processing, and amplifying units, like transistors in electronics. The time scale for termination of the rapid high-amplitude signals generated by such units, together with scales of other elementary cellular processes such as driving molecular motors and refractory periods of action potentials, determines the emergent scale of crucial life functions such as heartbeat, muscle motion, and brain activities.

As molecules interact, the signal termination is an emerging property of the channel cluster, rather than a property of an individual channel. This nanoscale system of interacting molecules which generates elementary biological signals is particularly interesting because this is where physics meets biology. Despite extensive experimental and numerical studies on the nanoscale, including structures of individual molecules and their locations (1-4), the universal links of synchronized intracellular signals to physics remain mainly unknown and conceptual understanding remains elusive.

The crucial importance of local Ca signaling amplified by Ca-induced-Ca-release (CICR) (5) for regulation of cardiac muscle contraction was theoretically predicted in 1992 (local control theory (6)), and further validated by the experimental discovery of Ca sparks (7). Sparks are generated by release channel clusters (Ca release Units, CRUs) in the form of localized Ca releases from junctional sarcoplasmic reticulum (SR). The sparks are strictly confined in time and space to roughly 20-40 ms by 2 μm by a powerful termination mechanism. Stochastic attrition, local SR depletion, coupled gating, channel inactivation, and lumenal regulation have been suggested as potential mechanisms of spark termination (review (8)). Since 2002 Sobie et al. model (9) the effect of mass action (decreasing Ca flux with Ca spark duration and loss of Ca in the SR) is a common driving element is virtually all of the spark models and central to the termination mechanism via interrupting CICR (10) in "pernicious attrition" (11) or "induction decay" phenomenon (12), i.e. the process opposed to CICR that facilitates channel closure within the CRU. The model formulations have improved further via superresolution modeling of calcium release (4), but the only "feature" that appears to have disappeared from all current models is the need for RyR inactivation. Thus the current consensus in the field on the spark termination problem is that SR depletion decreases the RyR channel current and therefore RyR interactions to the point that CICR can be sustained any longer, culminating in spark termination. At the same time, previous studies have not provided a mathematical understanding of this emerging behavior.

Here we apply statistical mechanics to describe quantitatively and deterministically the collective behavior of release channels within CRUs during spark termination. Specifically, we show that the channel synchronization that corresponds to termination of signals generated by a CRU is described by the same equation (and thus works on the same principle) as the phase transition associated to the change of magnetic field in an Ising ferromagnet.

The Ising model describes ferromagnetism using a square grid of interacting binary (+/-1) random variables representing atomic spin. An analytic solution of the 2D-case of this model was found by Onsager (13). The spins "want" to be aligned, and in the absence of magnetic field they can synchronize to either a +1 or -1 state. The application of a magnetic field breaks this symmetry, and the spins synchronize according to the sign of the magnetic field culminating in the phase transition. Thus the magnetic field indicates which state the spins "prefer." As described above, Ca release channels also "want" to be aligned in either closed or open state. An open release channel tries to open its neighbors via CICR (5), and similarly a closed release channel favors a closed state of its neighbors (12).

Following this analogy, we construct an exact mapping between two systems in 2 dimensions: a system of release channels within a CRU and the Ising model of interacting spins within a ferromagnet. Our mapping allows us to describe both systems by the same mathematical formulations. We take advantage of the extensive research on Ising models (since 1920 (14)); specifically, our criterion for spark termination in a CRU is the same as the well-known fact that the Ising model undergoes a phase transition when the external magnetic field reverses. This criterion closely predicts the depletion level required for spark termination for each CRU size. We demonstrate this mechanism using numerical model simulations of Ca sparks over a wide range of CRU sizes from 25 to 169 release channels.

**Results**

We linked a system of release channels within a CRU and the Ising model of interacting spins within a ferromagnet via the same mathematical formulations. Here we formulate four general conditions that a system of release channels should satisfy for our mapping between the two systems to hold. When these conditions are satisfied the RyR interactions are quantitatively identical to the interactions in a carefully constructed version of the Ising model (see Methods):

1. The channels are arranged in a lattice structure.
2. Each channel can be in an open or closed state, corresponding to the plus and minus states of the Ising model.
3. In the absence of interactions, the Ca profile resulting from one channel opening is roughly stable in time, corresponding to time-invariance of channel interactions (achieved e.g. by the joint action of buffer and diffusion out of the channel cluster), and roughly the same for any channel in the cluster, corresponding to translation invariance.

4. The channel opening rate as a function of Ca can be described by an exponential and the closing rate by a constant.

We vividly demonstrate the above principles by mapping a recent numerical model of sparks in cardiac cells (10) (referred hereafter as Stern model) (Fig. 1) to Ising model. Ryanodine Receptors (RyRs) in cardiac cells are indeed positioned in a crystal-like square grid and generate sparks by their synchronous opening (Movie S1 and S2 for different grid sizes). The closing rate of RyR is shown to be approximately constant in experimental studies (12) and it has been set to a constant in the original Stern model (10). RyR opening rate is described by an exponential $\lambda e^{\gamma [Ca]}$ (red line in Fig. 2A), and Ca profiles are roughly stable in time (Fig. 2B). Please note that the exponential in Figure 2A fits experimental data well only in our range of interest from 10 to 100 μM (spark termination range): beyond 100 μM the data points show saturation, but at the low resting Ca of 0.1 μM, the opening rate is expected to be substantially lower than that approximated by $\lambda$ (to match the known resting spark rate). To make sure that our Ising model operates within the realistic exponential fit, we estimated [Ca] at channels in closed states (Fig.1C, bottom panel) using numerical simulations and found that on average this [Ca] stays below 80 μM (i.e. within our fit in Fig. 2A).

We next read off the Ca profile $\psi$ that results from the opening of one RyR (Fig. 2C) and we let the interaction profile $\phi(r) = \psi(Ur)/\psi(U)$ in our version of the Ising model be a normalized and scaled version of $\psi$ (U is the distance between channels). The Ca diffusion out of the CRU corresponds to imposing a negative boundary condition. Using #4 we can construct a dynamic for the Ising model that satisfies **detailed balance**. This means that the Markov chain is reversible and implies convergence to the explicitly known equilibrium (Gibbs) measure given in equation 2 (Methods).

From our exact mapping we derive a formula for the analogue of the magnetic field $h$ (Methods), which reflects an imbalance in the probability and indicates whether release channels "prefer" an open or a closed state:

$$h = \frac{1}{2\beta}\ln\left(\frac{\lambda}{C}\right) + 2\pi \int_{s>.5} \phi(s)ds .$$

Here $\beta=\gamma\psi$ is analogue of inverse temperature in classical Ising model; $C=117$ s$^{-1}$ is the channel closing rate; $\lambda$ and $\gamma$ determine the opening rate as $\lambda e^{\gamma[Ca]}$ (Fig. 2A). The transition is expected at $h$ just above 0, equivalent to reversal of polarity in ferromagnetism. Because $C$, $\lambda$, $\gamma$ are constants, but $\phi$ (and $\psi$) and its integral over neighboring channels depend on the grid size and Ca$_{SR}$ ([Ca] in SR), we explored $h$ vs. these two parameters and found that $h$ reverses at lower Ca$_{SR}$ levels in larger grids (Fig. 2D).

We tested Ising model predictions in *in silico* experiments using Stern model, in which Ca$_{SR}$ was clamped at various levels. As Ca$_{SR}$ increases from 0.1 to 0.15 mM the system undergoes a sharp transition from terminating sparks to sparks that do not terminate (Fig. 3A). At higher Ca$_{SR}$ levels, the local [Ca] remains high, despite a negative boundary effect (Fig. 3B).

We further examined statistics of the termination time transition for different grid sizes and compared them to the Ising model prediction. As Ca$_{SR}$ increases, the termination time indeed undergoes a sharp increase (>10,000 ms) due to a phase transition for all grid sizes tested (Fig. 4A). The phase transition onsets are well-described by an exponential function (Fig. 4B) with steeper exponential rise for larger grids (Fig. 4C). The Ca$_{SR}$ at which $h$ reverses corresponds closely to the onset of phase transition for each grid size, specifically about 30% increase in termination time in our numerical simulations (Fig. 4D). This criterion remains effective for physiological conditions, when Ca$_{SR}$ is not fixed; namely, if a spark reaches the *h* reversal it terminates; otherwise it becomes metastable (Fig. 4E,F and Movies S3,S4).

**Discussion**

In the present study we approach the long-standing problem of calcium spark termination by an unexpected (for biologists and physiologists) application of statistical mechanics to the nanoscale biological signaling. We discovered that the spark termination process is mathematically isomorphic (governed by the same equations) to the phase transition associated with the reversal of magnetic field in a classical Ising ferromagnet. As this phase transition in ferromagnets is well-understood, we used this knowledge to formulate a criterion for the phase transition determining the spark termination in the CRU system. Our numerical model simulations have validated the theoretical prediction of the spark termination: they clearly show that sparks terminate when Ca SR depletion reaches the predicted level of the phase transition for each CRU size. Conversely, sparks do not terminate when SR depletion does not reach the phase transition level.

Abnormal sparks which don't promptly terminate (dubbed metastable sparks (10), embers (15), or "leaky" releases (16)) have been generated by numerical model simulations (10, 17, 18) and observed experimentally (19). In cardiac muscle, failure of Ca release to terminate leads to explosive Ca waves, which cause life-threatening arrhythmia (20, 21). Our finding of the release termination failure at higher Ca$_{SR}$ levels (Figs. 3,4, Movie S4) is in line with experimental observation of long-lasting Ca releases (19) and Ca waves (22) under Ca overload or rapid intra-SR Ca diffusion and re-uptake (10, 18), facilitating CICR among RyRs and CRUs. Recent numerical studies demonstrated that conditions favoring long-lasting sparks actually include a rather complex (and often counter-intuitive) interplay of SR Ca loading, number of functional RyRs, and RyR gating kinetics (10, 17, 18). These conditions can be now recast and explained in the new terms of the phase transition criterion formulated in the present study. Thus, our approach could be helpful in effective predicting and directing drug actions to avoid the metastable spark regime and to normalize cardiac rhythm.

One important immediate advantage the CRU to Ising mapping is that it provides a new, much more efficient computational means for evaluating the behavior of models of Ca release channel clusters. Indeed, while the Markov chain formulation has an analytic solution for its steady-state, the number of states is very large for a reasonable size cluster of release channels. For example, the number of states is 2^(number of channels) that is $2^{169} \sim 10^{51}$ (for a RyR cluster of 13 x 13). Thus, computing the dynamics for all states in the full Markovian representation using the analytic solution to Markov matrix involves taking exponentials of huge matrices, which is impractical. Moreover, the full CRU model also includes the dynamics of diffusion of Ca within the cleft, which is not explicitly represented in the Markov model. The only practical way to study such a complex system is by numerical simulations (10, 12) which still require many hours of computing on powerful computer clusters. The analytic representation using statistical mechanics suggested in the present study is much more compact and efficient. It allows the evaluation of the model behavior within milliseconds of computing time for a given channel interaction profile and CaSR level. For example, our rough evaluation of phase transitions for 5 grid sizes required 23,300 simulation runs (Fig. 4B and Fig. 4D, red symbols), while the same transitions were found virtually instantly by using equation 10 (Fig. 4D, blue symbols).

Our study extends the domain of applicability of statistical mechanics, which traditionally describes systems with numerous (or infinite) number of elements. Here we show that this relatively small biological system with as few as 25 molecules is able to function utilizing the smoothed-out phase transition associated with finite systems (Fig. 4, Movie S2). While larger clusters, such as of 169 molecules, generate stronger local signals, they are "harder" to terminate (i.e. lower $Ca_{SR}$ levels are required, Fig. 4A). Super-resolution imaging data on CRU ultrastructure show that actual RyR clusters in cardiac cells exhibit complex shapes and various sizes, with some CRUs being an incompletely filled grid of channels (3, 4) that could help reach a balance between signal strength and termination. In cardiac pacemaker cells, where diastolic Ca releases are synchronized via local propagation (23, 24), bigger RyR clusters are mixed with smaller "connecting" clusters (25). The Ising model with some sites missing would be a dilute Ising model, which is currently an object of active research in physics (26).

The equivalence of an Ising and CRU model can tell us the limiting probability of any configuration of open/closed release channels. It will be given by the equilibrium measure for the corresponding Ising model, and will depend only on the length of the contour. With this tool one can quantify the extent of termination failure in terms of the equilibrium measure. Such an application of interacting particle systems bridging the gap between scales of individual molecules and their collective behavior has been a long-standing problem in biology (27).

Ca puffs generated by IP3 receptors (IP3r) are generally accepted to be collective events in which clustered channels are mutually activated by CICR. Their termination mechanism, remains uncertain (28). Our present model cannot be directly applied to puffs, because unlike the RyR, the IP3r is strongly inactivated by high calcium, meaning that our condition (4) only holds for a small part of the calcium range. There is considerable uncertainty about the mechanisms of IP3r opening and closing in puffs (28-30), so it is premature to speculate on details of any possible extension of the Ising paradigm to puffs. On the other hand, in 2009 Smith and Parker (31) resolved individual openings and closings of IP3r's during a puff that showed a tendency for IP3r's to close abruptly and collectively ("square" puffs). Wiltgen et al. (28) examined further the "square puff" phenomenon and showed unambiguously that IP3r's in a decaying puff do not behave independently, but tend to close synchronously. This implies some kind of "closing signal" coordinating the various channels. The nature of this signal remains unknown. However, the collective inactivation may be suggestive of some kind of phase transition, and a "closing signal" could be accommodated in an Ising-like model.

**Methods**

1. **Brief Introduction to the Ising Model.** The Ising model we will work with consists of binary random variables (i.e. taking values ±1) called **spins** positioned on a 2D finite grid $\Lambda$ (e.g. section 3.3.5 in (32)). A configuration of spins is a function σ that assigns 1 or -1 to each point $x \in \Lambda$. The configuration space $\Omega$ is the set of all possible assignments of spins to points in $\Lambda$, i.e. all possible functions $\sigma : \Lambda \to \{1, -1\}$. A **interaction profile** $\phi: \mathbb{R} \to \mathbb{R}$ is a function with $\phi(x) \to 0$ rapidly as $x \to \infty$ and $\phi > 0$. We choose $\phi$ so that $\phi(1) = 1$. We furthermore place our finite grid $\Lambda$ inside of a bigger grid $\Lambda_b$ (b for boundary) and let $\sigma(x) = -1$ for any $x \in \Lambda_b \setminus \Lambda$. In this way we impose a -1 **boundary condition** on $\Lambda$. Here $\Lambda_b \setminus \Lambda$ must "frame" $\Lambda$ and its thickness has to be at least as wide as the effective interaction range, which in our case will be around 5. To be precise, if $\Lambda$ is a *n* by *m* grid, $\Lambda_b$ will be a *n+10* by *m+10* grid with $\Lambda$ situated in the middle of $\Lambda_b$. The Hamiltonian is

[1] $$H(\sigma) = - \sum_{x,y \in \Lambda_b} \phi(|x-y|)\sigma(x)\sigma(y) - h \sum_{x \in \Lambda_b} \sigma(x)$$

Here the first sum is over $\Lambda_b$ instead of $\Lambda$. This is necessary to ensure the interaction with the boundary.

In physics, *h* is the magnetic field. The Hamiltonian can be interpreted as the energy of the system. The equilibrium measure (Gibbs measure) is given by

[2] $$\pi(\sigma) = Z^{-1} e^{-\beta H(\sigma)}.$$

The normalization constant *Z* is well-defined since our lattice $\Lambda$ is finite, and we will not need to know it explicitly for our analysis. Here *β* is the inverse temperature. (For further information on the general Ising model, of which this is an instance, cf Sections 2.1 and 2.2 in (33))

2. **Dynamic Ising: Detailed Balance and the Transition Rates.** Let $\Lambda$ be a 2 dimensional integer grid of a finite size. Recall that $\Omega$ is the configuration space and let $\sigma: \Lambda \to \{1, -1\}$ be an element of $\Omega$. One can introduce a dynamic on spin configurations so that the configuration space $\Omega$ becomes the state space for a Markov chain with a transition matrix $P$. We introduce the notation $\sigma^x$ to mean

$$\sigma^x = \begin{cases} \sigma(y) & for\ y \neq x \\ -\sigma(y) & for\ y = x \end{cases}$$

i.e. $\sigma^x$ coincides with $\sigma$ everywhere except at $x$, where the spin is reversed. To obtain a Glauber-like dynamic for the Ising model, it suffices to choose a spin uniformly at random at each time increment and to give the probability that it flips, i.e. to give $P(\sigma \to \sigma^x)$.

The condition on $P$ that guarantees that $\pi$ as in [2] is indeed the equilibrium measure for the Markov chain is called detailed balance, and it states that the Markov chain is reversible with respect to $\pi$ (cf equation (1.30) and Proposition 1.19 in (32)). The equation for detailed balance is the following: for all $\sigma \in \Omega$ and $x \in \Lambda$ we have that

[3] $$P(\sigma \to \sigma^x)e^{-\beta H(\sigma)} = P(\sigma^x \to \sigma)e^{-\beta H(\sigma^x)}$$

This is equivalent to

[4] $$\frac{P(\sigma \to \sigma^x)}{P(\sigma^x \to \sigma)} = e^{\beta H(\sigma) - \beta H(\sigma^x)}$$

$$= e^{-2\beta(\sum_{y \in \Lambda_b} \phi(|x-y|)\sigma(x)\sigma(y) + h\sigma(x))} = e^{-2\beta\sigma(x)(\sum_{y \in \Lambda_b} \phi(|x-y|)\sigma(y) + h)}$$

The detailed balance equations will be satisfied for a wide variety of rates $P$, so we can choose $P$ to be most appropriate to our CRU model. Since we know that the release channel opening rate is an exponential while the closing rate is a constant, we look for $P$ so that the transition from -1 to 1 is exponential while the transition from 1 to -1 is a constant. This indeed can be achieved simultaneously with the detailed balance condition. If $\sigma(x) = -1$ we let

$$P(\sigma \to \sigma^x) = Ce^{2\beta\left(\sum_{y \in \Lambda_b} \phi(|x-y|)\sigma(y)+h\right)}$$ yielding that $P(\sigma^x \to \sigma) = C$ to satisfy detailed balance. Thus, the Markov chain is given as follows. We pick a location $x$ uniformly at random, and define the transition matrix $P$ to be:

[5] $$P(\sigma, \sigma^x) = \begin{cases} Ce^{2\beta(\sum_{y \in \Lambda_b} \phi(|x-y|)\sigma(y)+h)} & \text{for } \sigma(x) = -1 \\ C & \text{for } \sigma(x) = 1 \end{cases}$$

Here time is continuous and the above are transition rates. In our numerical model, time is discrete and we take $\Delta t = 0.05$ ms. The transition matrix with the discretized time becomes

[6] $$P(\sigma, \sigma^x) = \begin{cases} \Delta t Ce^{2\beta(\sum_{y \in \Lambda_b} \phi(|x-y|)\sigma(y)+h)} & \text{for } \sigma(x) = -1 \\ \Delta t C & \text{for } \sigma(x) = 1 \end{cases}$$

and we ensure that $\Delta t$ is small enough so that all transition probabilities are smaller than 1. Letting also $P(\sigma, \sigma) = 1 - P(\sigma, \sigma^x)$ ensures that $P$ is indeed stochastic.

**3. The CRU as an Ising Model.** A numerical model of the CRU consists of a square grid of calcium release channels $\Lambda$ and each release channel can be open or closed. We assign 1 to each open and -1 to each closed release channel, thus obtaining a configuration $\sigma : \Lambda \to \{1, -1\}$. We introduce the constant $U$ to represent the spatial distance between nearest release channels. In our numerical model, is $U = 30$ nm.

We let $\psi$ be the 1D slice of the time-stable spatial calcium profile resulting from the opening of one release channel. This is sufficient to contain all the information about the calcium profile since $\psi$ is rotationally symmetric. We obtain $\psi$ from our numerical simulation. However, $\psi$ is an immediate result of the environment, including current, diffusion, and buffer and is not an emergent property. We interpret it as a scaled interaction profile, and let $\phi$ in [1] be given as $\phi(r) = \psi(Ur)/\psi(U)$, where $Ur$ is the distance to the open release channel. The multiplication by *U* accounts for the fact that the release channels are *U* units apart while spins are 1 unit apart. The division by $\psi(U)$ is a choice of scaling for the interaction profile function $\phi$. With this scaling we have $\phi(1) = 1$. We choose this scaling for $\phi$ so that at the nearest neighbors its value matches the classic Ising model, where each spin interacts with 4 neighbors with a strength of 1.

The distance between CRUs is assumed to be too large for calcium from one CRU to influence another. On the other hand, calcium is diffusing out of the CRU and in this way the release channels in the CRU interact with the outside. The model would be identical if the CRU were surrounded by release channels that are always closed. In this way, the boundary condition of the CRU model is equivalent to a negative boundary condition of the Ising model.

We will compute the analogues of inverse temperature *β* and the magnetic field *h* in our CRU model as functions of initial model parameters. They play the exact same role in the mathematical description of our CRU model as they do in the Ising model even though they do not carry the same physical meaning. We will note that *β* is an increasing function of the concentration of Ca inside the junctional SR and we vary the SR Ca in our numerical model to test the predictions of the CRU Ising model.

**4. Relating [Ca] and the Ising Hamiltonian.** Let us introduce the set $S(x) := \{s \in \mathbb{R} : s = |x - y| \neq 0 \text{ for some } y \in \mathbb{Z}^2\}$. We can rewrite both the local [Ca] at *x* (we denote it [Ca](*x*)) and the exponent in the -1 to 1 transition in *P* in terms of a sum over $S(x)$. Given a configuration of open and closed release channels $\sigma$ and a given release channel at a point *x*, let $N_{Us}$ be the number of open RyRs at a distance *Us* from *x*. If the release channel at *x* is closed, we can approximate [Ca] at *x* by

[7] $$[Ca](x) = \sum_{s \in S(x)} \psi(Us) N_{Us}$$

We similarly rewrite P. We introduce the following notation: $T_s(x) :=$ total number of spins at distance s from $x$; $L_s(x) :=$ number of -1 spins at distance s from $x$; $N_s(x) :=$ number of +1 spins at distance s from $x$; and we have $N_s(x) + L_s(x) = T_s(x)$.

Henceforth in this section, let us fix a site $x \in \Lambda$ and suppress the dependence on $x$ in $T_s$, $L_s$, $N_s$, and S for ease of notation. Then we can rewrite the expression in the exponent of the Ising -1 to +1 transition probability in [5] in the following way:

[8]
$$\sum_{y \in \Lambda_b} \phi(|x-y|)\sigma(y) = \sum_{s \in S} \phi(s)(N_s - L_s) = \sum_{s \in S} \phi(s)(2N_s - T_s)$$
$$= 2\sum_{s \in S} \phi(s)N_s - \sum_{s \in S} \phi(s)T_s \approx 2\sum_{s \in S} \phi(s)N_s - 2\pi \int_{s>.5} \phi(s)\,ds$$

In the last approximate equality, we have replaced $\sum_{s \in S} \phi(s)T_s$ by $2\pi \int_{s>.5} \phi(s)ds$ where the factor of $2\pi$ is due to the fact that $\sum_{s \in S} \phi(s)T_s$ is approximately a 2D integral of a rotationally symmetric function. We observe that the first term in the final expression in [8] is a scalar multiple of the total calcium [Ca] $(x)$ as given in [7].

**5. Crucial Parameters and the Spark Termination Criterion.** We want to solve for the analogues of $h$ and $\beta$ in the CRU model. We again fix a site $x \in \Lambda$ and suppress the dependence on $x$ in [Ca] and S for ease of notation. From experimental data we fit the exponential $\lambda e^{\gamma[Ca]}$ to the Ising transition rate from -1 to +1 in [5]:

$$\lambda e^{\gamma[Ca]} = Ce^{2\beta(\sum_{y \in \Lambda_b} \phi(|x-y|)\sigma(y)+h)}$$

Then we replace the LHS using [7] and the RHS using the expression derived in [8] to obtain

[9]
$$\lambda e^{\gamma \sum_{s \in S} \psi(Us)N_{Us}} = Ce^{2\beta(2\sum_{s \in S} \phi(s)N_s - 2\pi \int_{s>.5} \phi(s)\,ds+h)}$$
$$= Ce^{-4\beta\pi \int_{s>.5} \phi(s)+2\beta h} e^{4\beta(\sum_{s \in S} \phi(s)N_s)}$$

Since we wish the above equality to hold for any configuration, we must equate the coefficients of $\sum_{s \in S} \phi(s)N_s$ to obtain $\beta = \gamma\psi(U)/4$.

Next we equate the coefficients in front of $e^{4\beta(\sum_{s \in S}\phi(s)N_s)}$ to obtain

$$\lambda = Ce^{-4\beta\pi\int_{s>.5}\phi(s)ds + 2\beta h}$$

yielding that

[10] $$h = \frac{1}{2\beta}\ln\left(\frac{\lambda}{C}\right) + 2\pi\int_{s>.5}\phi(s)ds$$

Rewriting $h$ in terms of the calcium profile $\psi$ we obtain

[11] $$h = \frac{2}{\gamma\psi(U)}\ln\left(\frac{\lambda}{C}\right) + 2\pi\int_{s>U/2}\frac{\psi(s)}{U\psi(U)}ds$$

Since $h$ is the analogue of the magnetic field in the CRU model, the emergent behavior of release channels can be predicted based on $h$. During termination all the release channels begin in an open state (analogous to +1). The Ca diffusion out of CRU is equivalent to a negative boundary condition. We can hence deduce the **signal termination criterion**: If $h < 0$, then the spark will terminate and this termination is mathematically identical to reversal of polarity in ferromagnetism. Mathematically, this phase transition follows from the Lee-Yang theorem. On the other hand, if $h > 0$, the spark will be metastable.

**Acknowledgments**

The work was supported by the Intramural Research Program of the National Institute on Aging, National Institute of Health. AM acknowledges the support of the Leverhulme Trust Early Career Fellowship (ECF 2013-613).

**References**

1. Peng W, Shen H, Wu J, Guo W, Pan X, Wang R, Chen SR, & Yan N (2016) Structural basis for the gating mechanism of the type 2 ryanodine receptor RyR2. *Science* 354(6310).
2. Soeller C, Crossman D, Gilbert R, & Cannell MB (2007) Analysis of ryanodine receptor clusters in rat and human cardiac myocytes. *Proc Natl Acad Sci U S A* 104(38):14958-14963.
3. Baddeley D, Jayasinghe ID, Lam L, Rossberger S, Cannell MB, & Soeller C (2009) Optical single-channel resolution imaging of the ryanodine receptor distribution in rat cardiac myocytes. *Proc Natl Acad Sci U S A* 106(52):22275-22280.
4. Walker MA, Williams GS, Kohl T, Lehnart SE, Jafri MS, Greenstein JL, Lederer WJ, & Winslow RL (2014) Superresolution modeling of calcium release in the heart. *Biophys J* 107(12):3018-3029.
5. Fabiato A (1983) Calcium-induced release of calcium from the cardiac sarcoplasmic reticulum. *Am J Physiol* 245(1):C1-14.
6. Stern MD (1992) Theory of excitation-contraction coupling in cardiac muscle. *Biophys J* 63(2):497-517.

7.  Cheng H, Lederer WJ, & Cannell MB (1993) Calcium sparks: elementary events underlying excitation-contraction coupling in heart muscle. *Science* 262(5134):740-744.

8.  Cheng H & Lederer WJ (2008) Calcium sparks. *Physiol Rev* 88(4):1491-1545.

9.  Sobie EA, Dilly KW, dos Santos Cruz J, Lederer WJ, & Jafri MS (2002) Termination of cardiac Ca(2+) sparks: an investigative mathematical model of calcium-induced calcium release. *Biophys J* 83(1):59-78.

10. Stern MD, Rios E, & Maltsev VA (2013) Life and death of a cardiac calcium spark. *J Gen Physiol* 142(3):257-274.

11. Gillespie D & Fill M (2013) Pernicious attrition and inter-RyR2 CICR current control in cardiac muscle. *J Mol Cell Cardiol* 58:53-58.

12. Laver DR, Kong CH, Imtiaz MS, & Cannell MB (2013) Termination of calcium-induced calcium release by induction decay: an emergent property of stochastic channel gating and molecular scale architecture. *J Mol Cell Cardiol* 54:98-100.

13. Onsager L (1944) Crystal Statistics. I. A Two-Dimensional Model with an Order-Disorder Transition. *Physical Review* 65:117-149.

14. Lenz W (1920) Beiträge zum Verständnis der magnetischen Eigenschaften in festen Körpern. *Physikalische Zeitschrift* 21:613–615.

15. Csernoch L (2007) Sparks and embers of skeletal muscle: the exciting events of contractile activation. *Pflugers Arch* 454(6):869-878.

16. Lehnart SE, Mongillo M, Bellinger A, Lindegger N, Chen BX, Hsueh W, Reiken S, Wronska A, Drew LJ, Ward CW, Lederer WJ, Kass RS, Morley G, & Marks AR (2008) Leaky $Ca^{2+}$ release channel/ryanodine receptor 2 causes seizures and sudden cardiac death in mice. *J Clin Invest* 118(6):2230-2245.

17. Song Z, Karma A, Weiss JN, & Qu Z (2016) Long-Lasting Sparks: Multi-Metastability and Release Competition in the Calcium Release Unit Network. *PLoS Comput Biol* 12(1):e1004671.

18. Sato D, Shannon TR, & Bers DM (2016) Sarcoplasmic Reticulum Structure and Functional Properties that Promote Long-Lasting Calcium Sparks. *Biophys J* 110(2):382-390.

19. Zima AV, Picht E, Bers DM, & Blatter LA (2008) Partial inhibition of sarcoplasmic reticulum ca release evokes long-lasting ca release events in ventricular myocytes: role of luminal ca in termination of ca release. *Biophys J* 94(5):1867-1879.

20. Sobie EA & Lederer WJ (2012) Dynamic local changes in sarcoplasmic reticulum calcium: physiological and pathophysiological roles. *J Mol Cell Cardiol* 52(2):304-311.

21. Qu Z & Weiss JN (2015) Mechanisms of ventricular arrhythmias: from molecular fluctuations to electrical turbulence. *Annu Rev Physiol* 77:29-55.

22. Ter Keurs HE & Boyden PA (2007) Calcium and arrhythmogenesis. *Physiol Rev* 87(2):457-506.

23. Maltsev AV, Maltsev VA, Mikheev M, Maltseva LA, Sirenko SG, Lakatta EG, & Stern MD (2011) Synchronization of stochastic $Ca^{2+}$ release units creates a rhythmic $Ca^{2+}$ clock in cardiac pacemaker cells. *Biophys J* 100:271-283.

24. Maltsev AV, Yaniv Y, Stern MD, Lakatta EG, & Maltsev VA (2013) RyR-NCX-SERCA local crosstalk ensures pacemaker cell function at rest and during the fight-or-flight reflex. *Circ Res* 113(10):e94-e100.

25. Stern MD, Maltseva LA, Juhaszova M, Sollott SJ, Lakatta EG, & Maltsev VA (2014) Hierarchical clustering of ryanodine receptors enables emergence of a calcium clock in sinoatrial node cells. *J Gen Physiol* 143(5):577-604.

26. Bodineau T, Graham B, & Wouts M (2013) Metastability in the dilute Ising model. *Probability Theory and Related Fields* 157(3):955-1009.

27. Qu Z, Garfinkel A, Weiss JN, & Nivala M (2011) Multi-scale modeling in biology: how to bridge the gaps between scales? *Prog Biophys Mol Biol* 107(1):21-31.

28. Wiltgen SM, Dickinson GD, Swaminathan D, & Parker I (2014) Termination of calcium puffs and coupled closings of inositol trisphosphate receptor channels. *Cell Calcium* 56(3):157-168.

29. Shuai J, Pearson JE, Foskett JK, Mak DO, & Parker I (2007) A kinetic model of single and clustered IP3 receptors in the absence of $Ca^{2+}$ feedback. *Biophys J* 93(4):1151-1162.

30. Ullah G, Parker I, Mak DO, & Pearson JE (2012) Multi-scale data-driven modeling and observation of calcium puffs. *Cell Calcium* 52(2):152-160.

31. Smith IF & Parker I (2009) Imaging the quantal substructure of single IP3R channel activity during $Ca^{2+}$ puffs in intact mammalian cells. *Proc Natl Acad Sci U S A* 106(15):6404-6409.

32. Levin DA, Peres Y, & Wilmer EL (2008) *Markov Chains and Mixing Times* ( American Mathematical Society) 1 Ed.

33. Martinelli F (1999) Lectures on Glauber Dynamics for Discrete Spin Models. *Lectures on Probability Theory and Statistics*, ed Bernard P (Springer-Verlag, Berlin Heidelberg).

**Figure Legends**

**Fig. 1: Stern numerical model describes collective behavior of RyR ensemble during spark activation and termination.** **A**, A schematic representation of a CRU in cardiac cells. **B**, An example of Ca spark generated by a CRU featuring 9x9 RyRs separated by 30 nm. The sequence of the RyR ensemble states is shown along with their instant local [Ca] on a grid with 10x10 nm computational voxels in the cleft. White up-arrows indicate open channels. Green down-arrows indicate closed channels (see also Movie S1 and S2, for different grid sizes). [Ca] is coded by red shades, saturated at 30 µM. **C**, Dynamics of open number of RyRs and SR [Ca] ($Ca_{SR}$) during the spark. Termination span of ~40 ms is shown by a blue shadow. Bottom panel: at each time sample, we collected information about [Ca] at all closed channels (i.e. ready to open) and report here respective maximum, mean, and minimum values.

**Fig. 2. Construction of an exact mapping between a CRU described by Stern model and the Ising model of interacting spins.** **A**, The exponential relation of RyR opening rate vs. [Ca] in the cleft. All previous models fit a power function to original data obtained in lipid bilayers. Here we fit an exponential (red line) to the same data points (original data and power fit are reproduced from Laver et al. (12) with permission). Thus, we replaced the quadratic opening rate in original Stern model with the exponential opening rate from this fit. **B**, Representative [Ca](t) when one RyR is open in the center of the grid: at the open RyR and its closest neighbor. **C**, A steady-state spatial [Ca] profile when one RyR is open in the center of 9x9 grid (similar profiles for other grid sizes are not shown). **D**, Plots of *h* as a function of $Ca_{SR}$ for 5 different grid sizes.

Inset: Ising model predicts phase transition as *h* reverses at different Ca$_{SR}$ for each grid size; the transition requires lower Ca$_{SR}$ levelsfor larger grids.

**Fig. 3. *In silico* Ca$_{SR}$ clamp experiments validating Ising model predictions for spark termination.** **A**, Evolution of 9x9 RyR ensemble at various Ca$_{SR}$ levels after all RyRs are set in the open state at time 0. Sparks do not terminate at Ca$_{SR}$ above 0.15 mM. The sharp transition in the numerical model behavior is in line with the Ising model prediction of the phase transition above 0.12 mM for the 9x9 grid. **B**, [Ca] profiles in the dyadic cleft for terminated and non-terminated sparks at Ca$_{SR}$ levels higher and lower the phase transition. [Ca] is coded by red shades, saturating at 50 µM.

**Fig. 4. Results of statistical analysis of our *in silico* experiments testing Ising model predictions for various grid sizes.** **A**, Median termination times (T$_t$) plotted as a function of clamped Ca$_{SR}$. Each data point was obtained from 100 simulation runs. **B**, The data set of panel A, but at a smaller scale. The transition onsets are closely described by an exponential (shown at the plot). **C**, The transitions are sharper for larger grids as occurring within smaller ranges of Ca$_{SR}$ (ΔCa$_{SR}$). **D**, *h* reversal (Ca$_{SR}$_h reversal) in our Ising model closely predicts the onset of the phase transition in our model simulations. The transition onset in the simulations is estimated as a 30% increase in the median T$_t$ (Ca$_{SR}$_30%_T$_t$_increase) calculated using respective exponential fits in panel B. **E and F**, Open RyRs and Ca$_{SR}$ dynamics in representative examples of stable and metastable sparks (13x13 RyR grid). The metastable spark was generated by increasing SR Ca refiling rate (TAUFILL was decreased from 6.5 ms to 1.5 ms). Inset shows a narrow margin for Ca$_{SR}$ that determines spark termination fate (see Movies S3 and S4 for more details).

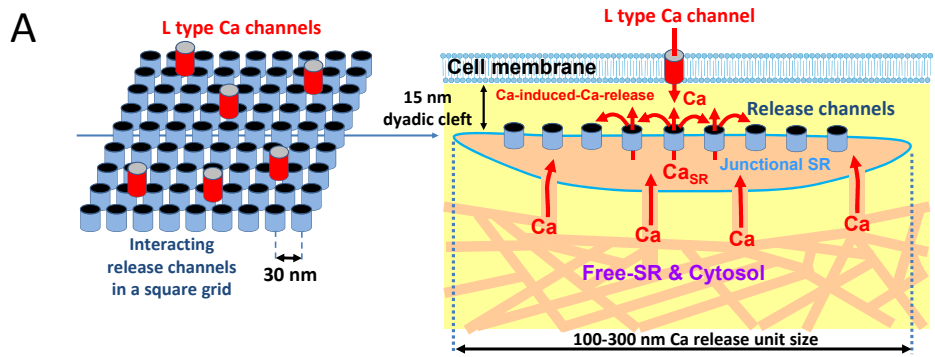

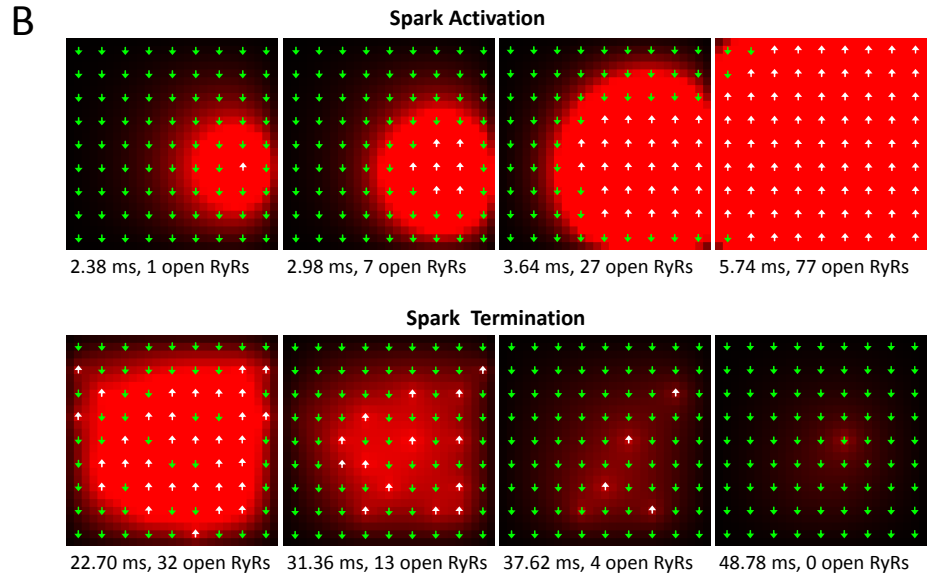

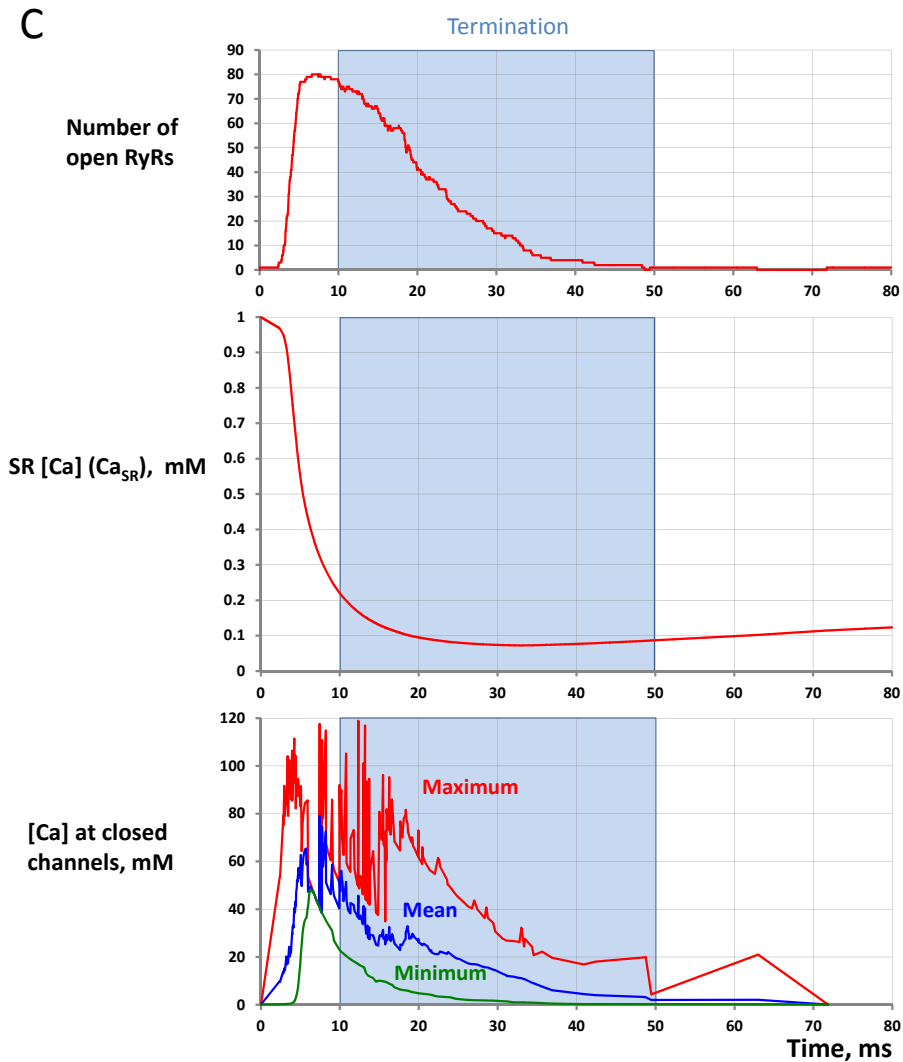

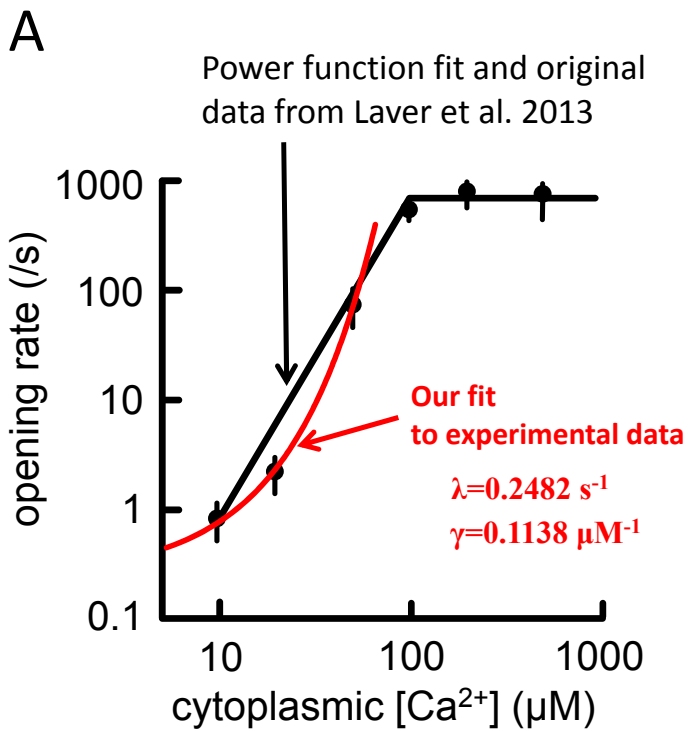
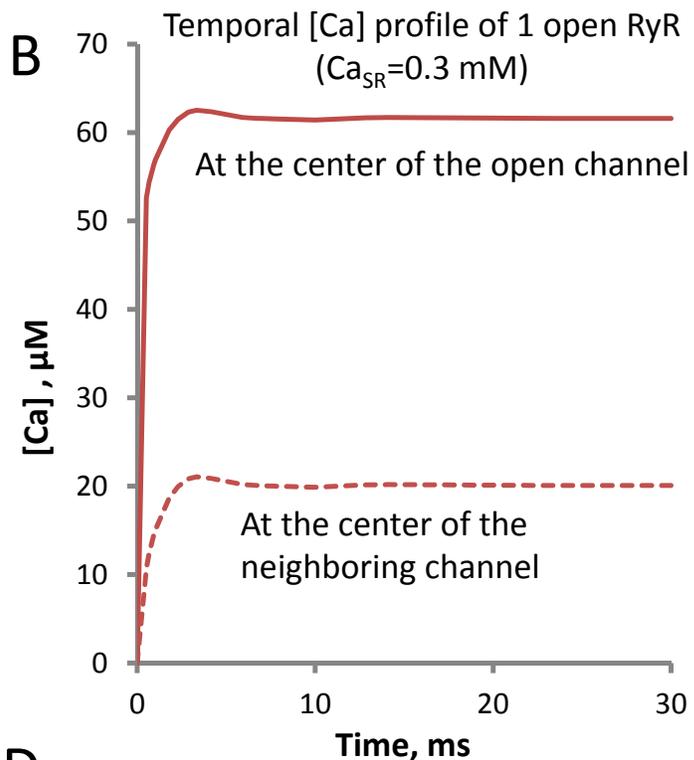
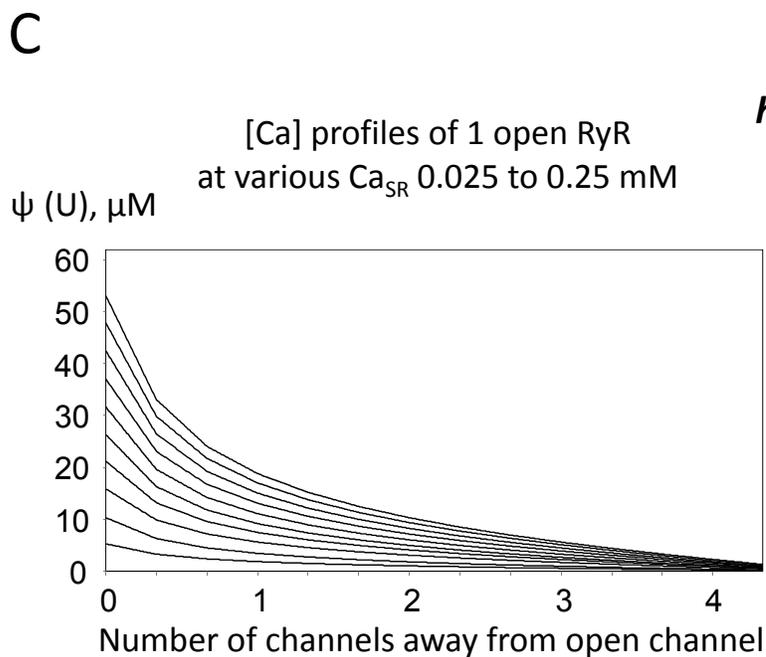
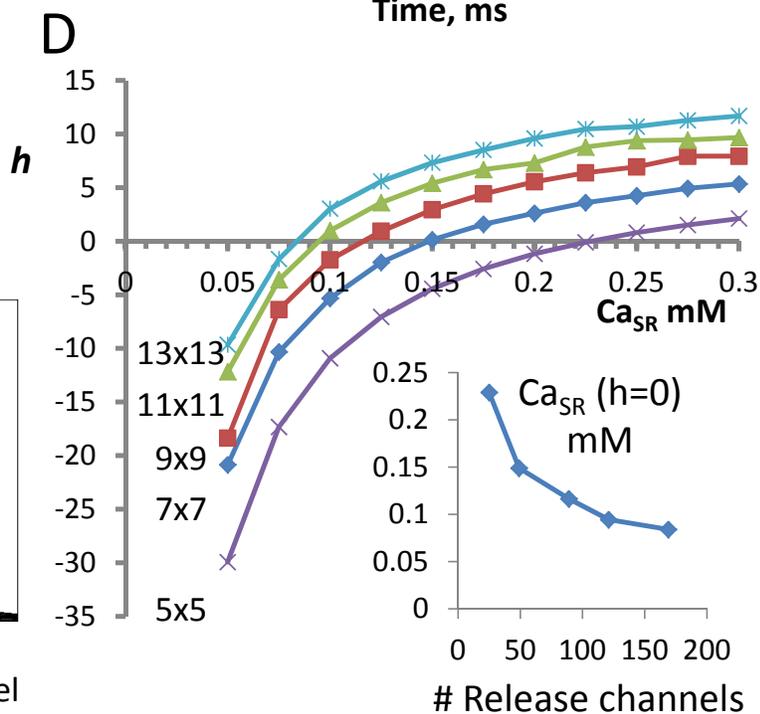

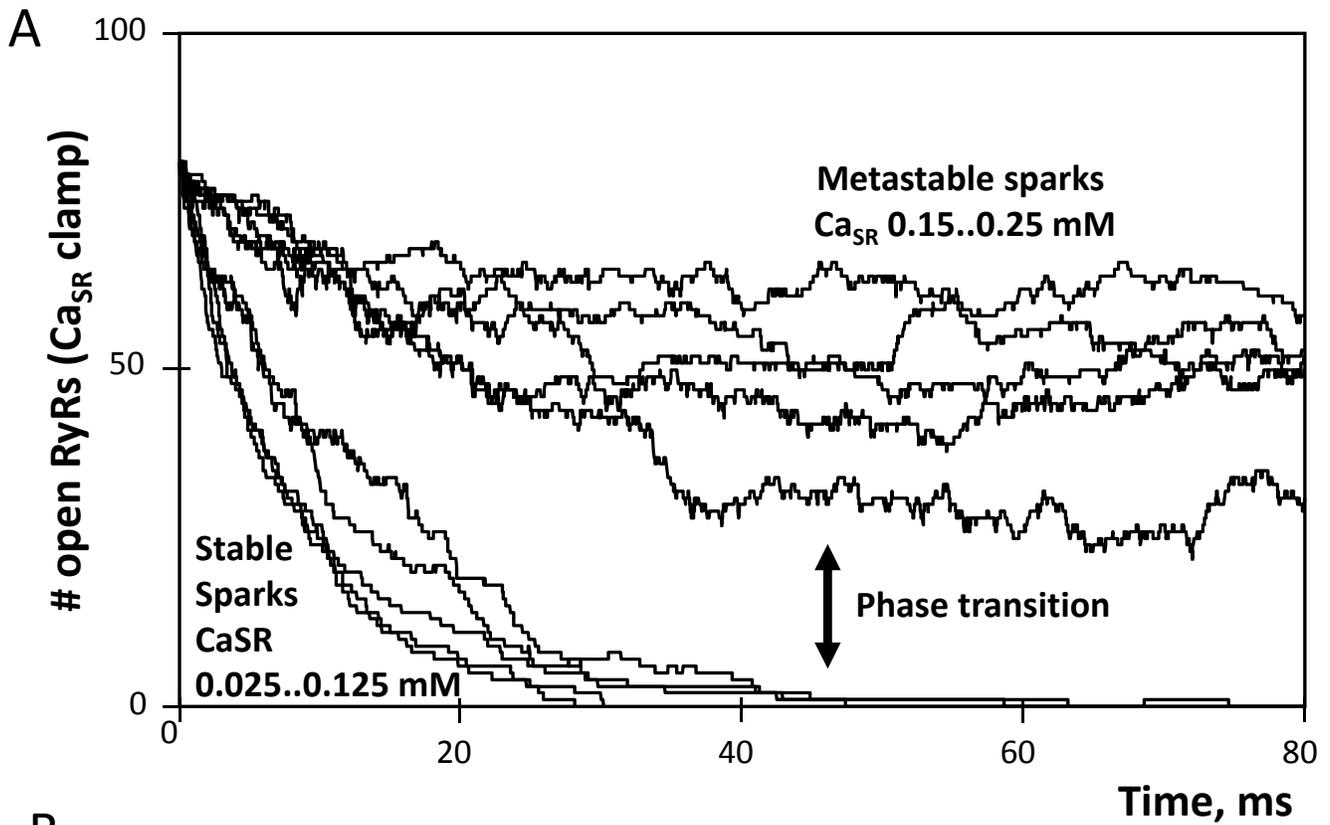

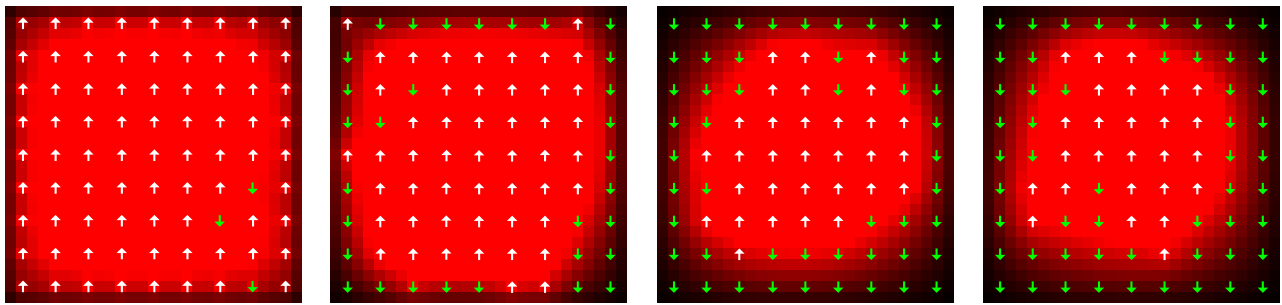

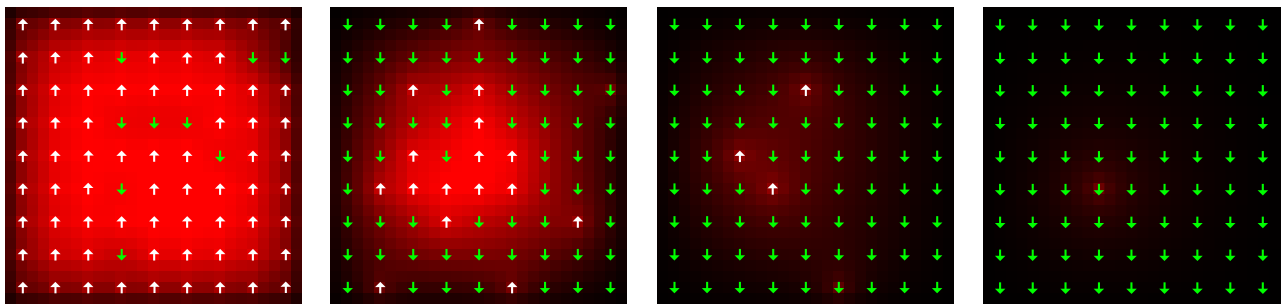

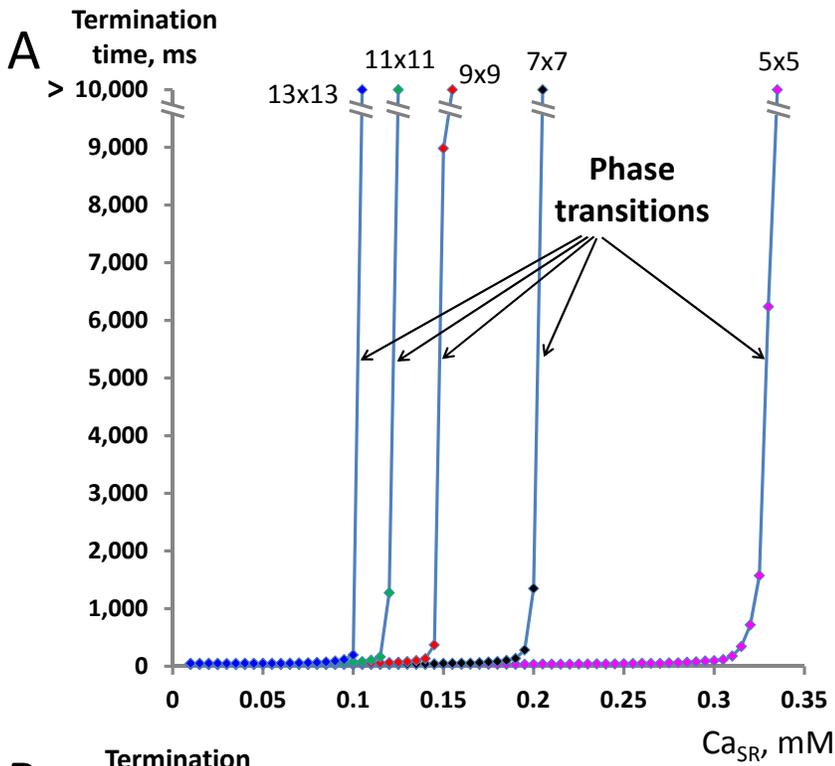
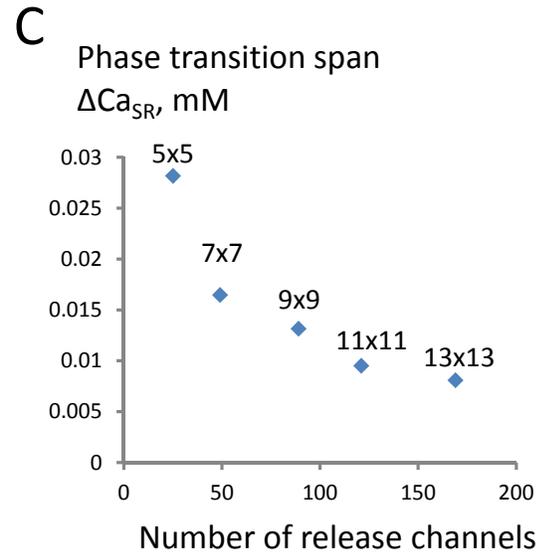
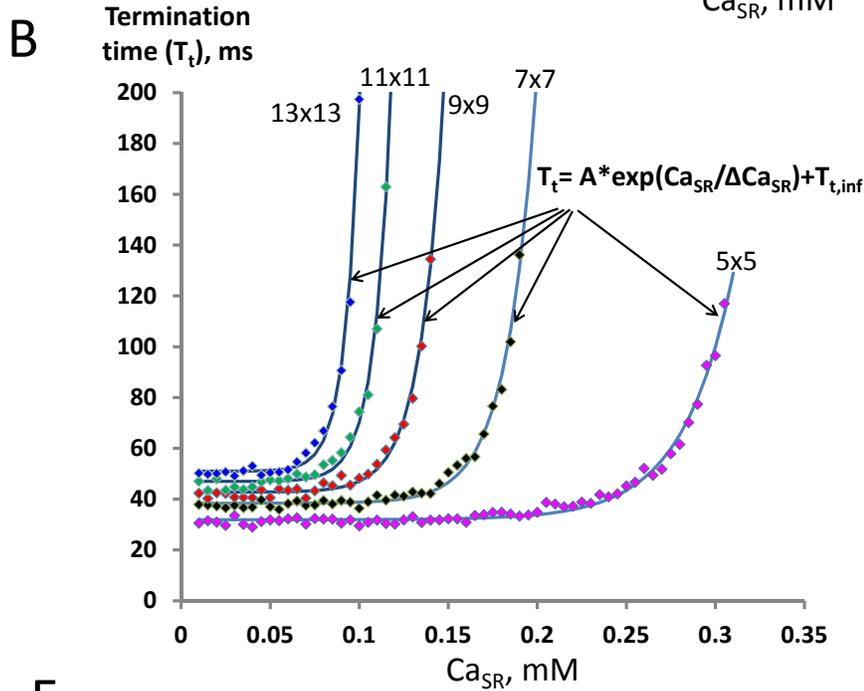
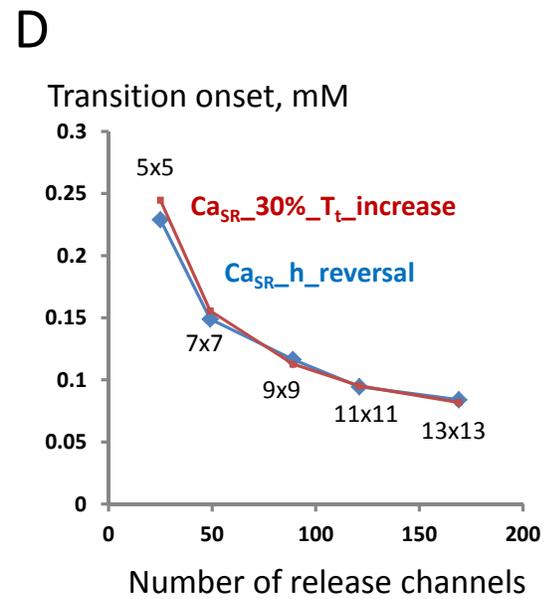
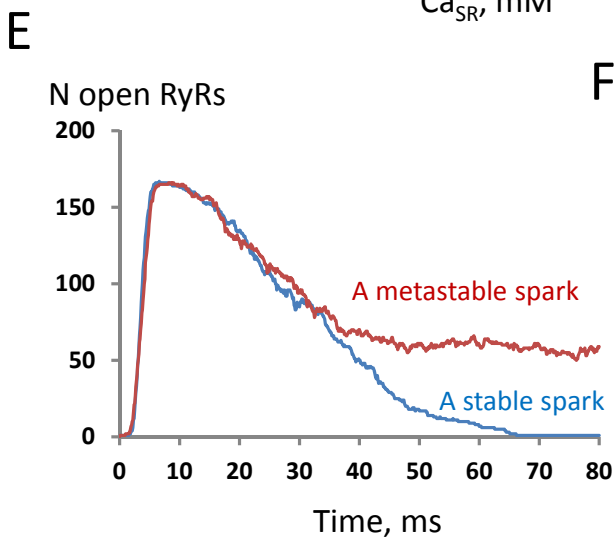
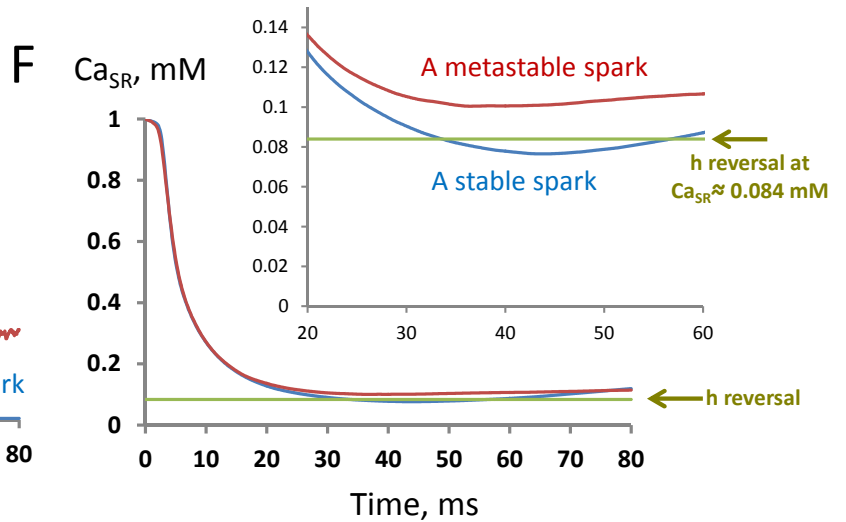

**Supporting Information: Movies**

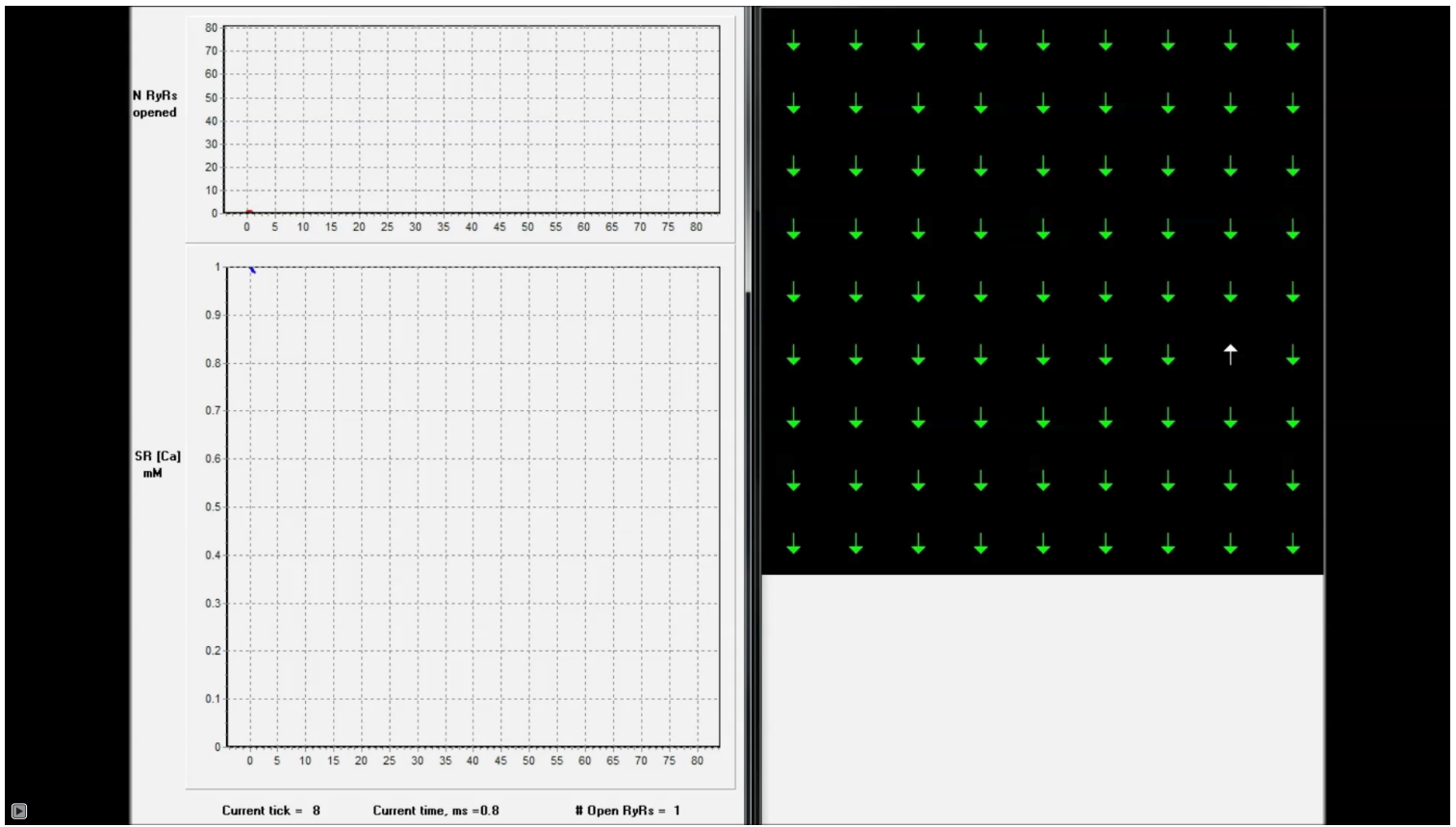

Link to Movie S1 in **mp4 format**.          Link to Movie S1 in **wmv format**

**Movie S1:** A typical Ca spark generated by Stern model by a mid-size release unit featuring 9x9 Ca release channels, ryanodine receptors (RyRs, marked by arrows, separated by 30 nm). Local [Ca] dynamics is simulated on a grid with 10x10 nm computational voxels in the dyadic cleft of 15 nm. White up-arrows indicate open channels. Green down-arrows indicate closed channels. [Ca] is coded by red shades, with pure red representing 30 μM. Left hand panels show the dynamics of the key spark parameters: open number of RyRs and SR [Ca] ($Ca_{SR}$).

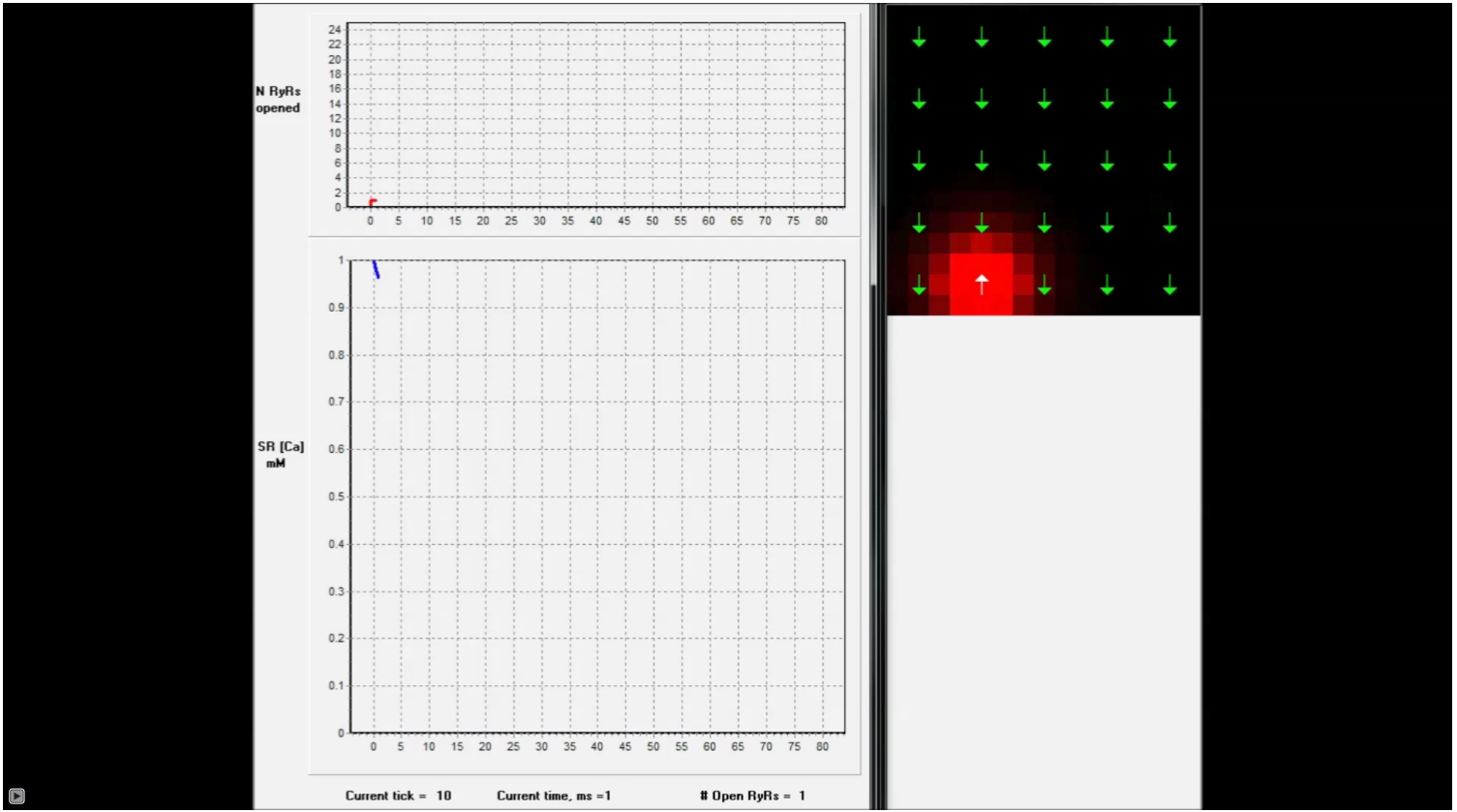

Link to Movie S2 in **mp4 format**.     Link to Movie S2 in **wmv format**

**Movie S2:** A spark generated by a very small release unit featuring 5x5 RyRs. RyR spacing, computational voxels, and [Ca] scale were 30 nm, 10x10x15 nm, and 30 μM, respectively, similar to Movie S1.

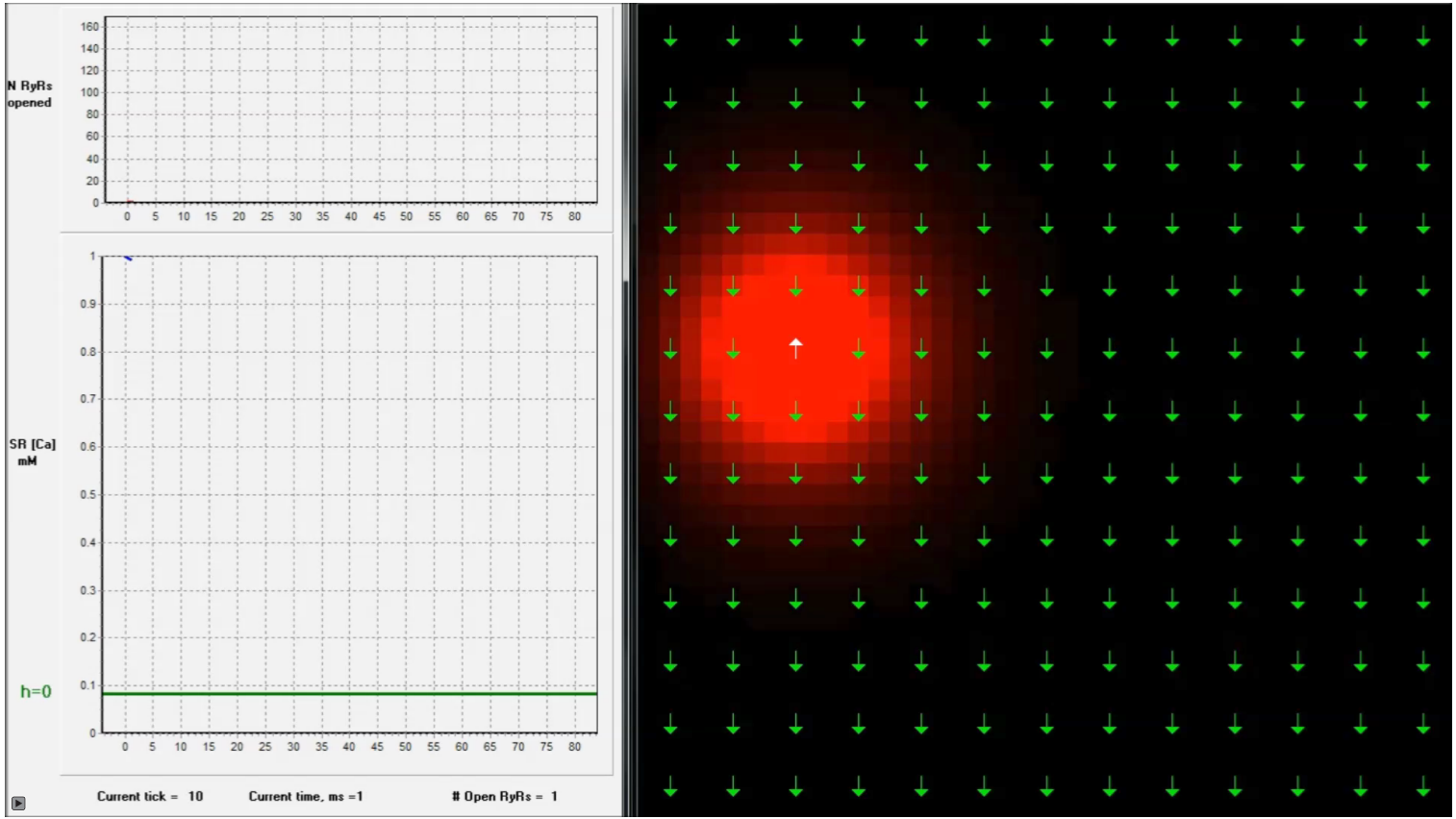

Link to Movie S3 in **mp4 format**.    Link to Movie S3 in **wmv format**

**Movie S3:** A stable spark generated by a large release unit featuring 13x13 RyRs. Green line shows Ca$_{SR}$ level of our Ising model prediction for phase transition (at h=0). RyR spacing, computational voxels, and [Ca] scale were 30 nm, 10x10x15 nm, and 30 μM, respectively.

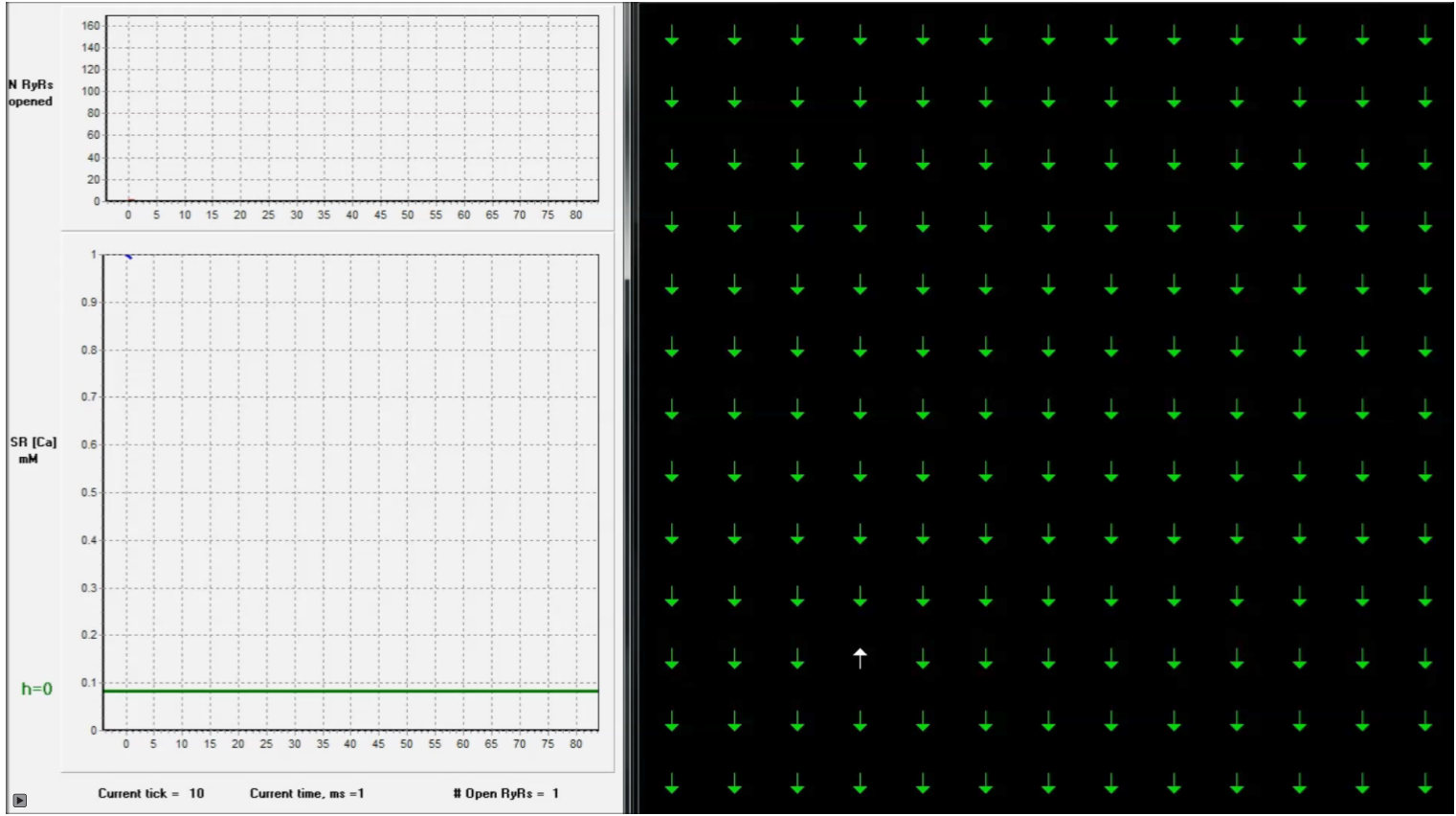

Link to Movie S4 in **mp4 format**.    Link to Movie S4 in **wmv format**

**Movie S4:** A metastable spark generated by a large release unit featuring 13x13 RyRs and increased junctional SR Ca refiling rate (TAUFILL was decreased from its original value of 6.5 ms to 1.5 ms). Green line shows Ca$_{SR}$ level of our Ising model prediction for phase transition (at h=0). RyR spacing, computational voxels, and [Ca] scale were 30 nm, 10x10x15 nm, and 30 μM, respectively.